\documentclass[12pt,a4paper]{article}

\usepackage[english]{babel}
\usepackage{amsmath,amsfonts,amssymb,amsthm}
\numberwithin{equation}{section}
\theoremstyle{plain}
\newtheorem{theorem}{Theorem}
\newtheorem{proposition}[theorem]{Proposition}

\newtheorem{lemma}[theorem]{Lemma}
\DeclareMathOperator{\ve}{\epsilon}
\newcommand{\bx}{\mathbf{x}}
\newcommand{\by}{\mathbf{y}}
\newcommand{\bz}{\mathbf{z}}
\newcommand{\ba}{\mathbf{a}}
\newcommand{\bu}{\mathbf{u}}
\newcommand{\bc}{\mathbf{c}}

\newcommand{\bR}{\mathbb{R}}
\newcommand{\bZ}{\mathbb{Z}}
\newcommand{\bN}{\mathbb{N}}

\begin{document}

\title{On the smoothness of gap boundaries for generalized Harper
  operators}

\author{Gheorghe Nenciu\\ 
Dept.  Theor. Phys., Univ. of Bucharest\\ P.O. Box MG 11, RO-077125, 
Bucharest, Romania\\
and \\ 
Institute of Mathematics of the Romanian Academy\\  
PO Box 1-764, 
RO-014700 Bucharest, Romania. \\
E-mail Gheorghe.Nenciu@imar.ro}
\date{}
\maketitle

\begin{abstract}
Results concerning set theoretic continuity properties of the spectrum of the
Harper operator are extended to a large class (generalized Harper operators
(GHO)) of operators in $L^{2}(\bZ^{2})$. 
\end{abstract}

\section{Introduction: the setting and the main results.}
Consider in $L^{2}(\bZ^{2})$ the following class of operators
\begin{equation}\label{model}
(h_{\ve})\psi(\bx)=\sum_{\by \in \bZ^{2}}e^{i\ve \phi(\bx,\by)}h(\bx, \by)\psi
(\by )
\end{equation}
where $\ve \in \mathbb{R}$ and $h(\bx, \by), \; \phi (\bx,\by)$ satisfy the
conditions:
\begin{equation}\label{herm}
h(\bx,\by)=\overline{h(\by,\bx)}
\end{equation}
\begin{equation}\label{marg}
|h(\bx,\by)|\leq Ce^{-\beta |\bx-\by|};\;\; C<\infty,\;\;0<\beta \leq 1
\end{equation}
\begin{equation}\label{antisim}
\phi (\bx, \by)= \overline{\phi (\bx, \by)}= -\phi (\by, \bx)
\end{equation}
\begin{equation}\label{flux}
|F(\bx, \by, \bz)|\leq area \;\;\Delta(\bx,\by,\bz)
\end{equation}
where
\begin{equation}\label{flux1}
F(\bx,\by,\bz)=\phi (\bx, \by)+\phi (\by, \bz)+\phi (\bz, \bx)
\end{equation}
and $\Delta(\bx,\by,\bz)$ is the triangle in $\mathbb{R}^{2}$ determined by
the points $\bx, \by, \bz $.

Under the conditions (\ref{herm}-\ref{flux1}), $h_{\ve}$ is a uniformly
bounded family of self-adjoint operators in $L^{2}(\bZ^{2})$.
While the family $h_{\ve}$ is (by a simple argument based on Lebesgue
dominated convergence theorem ) strongly continuous, in general it is not
normic continuous and this makes the problem of set theoretic continuity
properties of the spectrum, $\sigma (h_{\ve})$, a highly nontrivial one.

Some particular cases of the above class (called for the reasons below)
generalized Harper operators (GHO)) are well known. If
\begin{equation}\label{trans}
h(\bx,\by)=h(\bx -\by),
\end{equation}
\begin{equation}\label{const}
\phi(\bx,\by)=-x_{1}y_{2}+x_{2}y_{1};\;\; \bx =(x_{1},x_{2}),\;\by
=(y_{1},y_{2}),
\end{equation}
then $h_{\ve}$ are nothing but the "magnetic matrices" appearing in so called
Peierls-Onsager substitution for 2D-electrons moving in a periodic potential
and subjected to a constant magnetic field \cite{nen1},\cite{nen2},\cite{hesj}.
Also, they are the discrete version of twisted convolutions \cite{stein}.
If, in addition
\begin{equation}
h({\bf c})=\left\{\begin{array}{rcl}h& \mbox{if}& |{\bf x}| =1\\ 
0& othervise \end{array}\right. 
\end{equation} 
one recovers the famous Harper operator which, as well known, has a
fascinating Hofstatder butterfly like spectrum (see e.g. \cite{hegui},
\cite{boca} and references therein). In particular it has been conjectured that
for all irrational $\ve$, the spectrum has a Cantor set structure. The proof
of this conjecture for some classes for irrational $\ve$ \cite{ellchyu} used
in an essential way set continuity properties of $\sigma(h_{\ve})$.

The aim of this note is to prove that most of the set theoretic continuity
properties of $\sigma(h_{\ve})$ known for the particular case given by
(\ref{trans}),(\ref{const}) hold true in the general case. More precisely:
\begin{theorem}\label{thm1/2}
For
$h_{\ve}$  given by (\ref{model})-\ref{flux1}):

i. Let $E \in \sigma(h_{\ve})$. Then there exists an absolute constant $K
<\infty$, such that for $|\ve-\eta|\leq 1/2$
\begin{equation}\label{1/2}
dist(E, \sigma(h_{\eta})) \leq K\frac{C}{\beta^{6}}|\ve -\eta|^{1/2}.
\end{equation}
ii. Let
\begin{equation}
E_{+}(\ve)=sup \;\sigma(h_{\ve}).
\end{equation}
Then there exists an absolute constant $K<\infty$, such that for
$|\ve-\eta|\leq 1/2 $:
\begin{equation}\label{ln0}
|E_{+}(\ve)-E_{+}(\eta)| \leq K\frac{C}{\beta^{4}}|\ve -\eta||ln|\ve-\eta||.
\end{equation}
\end{theorem}
Suppose now that for some $\ve_{0}$ there is a gap $\Delta=(a,b),\;\;b-a=4d>0$
in the spectrum of $h_{\ve_{0}}$, i.e.
$\sigma(h_{\ve_{0}})=\sigma_{1}(\ve_{0})\cup \sigma_{2}(\ve_{0})$,
$sup\sigma_{1}(\ve_{0})=a,\;inf\sigma_{2}(\ve_{0})=b$. Then from Theorem 1i.
one has that for $|\ve-\ve_{0}|\leq nd^{2}$, where $n>0$ is a constant depending
upon $C$ and $\beta$, $h_{\ve}$ still has a gap, $\Delta(\ve)$ of length
larger than $2d$, i.e.
$\sigma(h_{\ve})=\sigma_{1}(\ve)\cup \sigma_{2}(\ve)$,
$\inf_{E\in \sigma_{1}(\ve),E'\in \sigma_{2}(\ve)}|E-E'|\geq 2d$ and
$\sigma_{1}(\ve)$ coincide with $\sigma_{1}(\ve_{0})$ in the limit $\ve
\rightarrow \ve_{0}$. So for $|\ve-\ve_{0}|\leq nd^{2}$ one can define
\begin{equation}
E_{1}(\ve)=sup\sigma_{1}(\ve),\;\;E_{2}(\ve)=inf\sigma_{2}(\ve).
\end{equation}
\begin{theorem} 
There exist constants $m>0,\;\;M<\infty$, depending upon $C$ and $\beta$ such
that for
\begin{equation}
|\ve-\ve_{0}|\leq md^{4},
\end{equation}
\begin{equation}\label{ln1}
|E_{j}(\ve)-E_{j}(\ve_{0})|\leq
M|\ve-\ve_{0}|(d^{-7}+d^{-5}|ln|\ve-\ve_{0}||).
\end{equation}
\end{theorem}
Theorem 1i. implies that for $|\ve-\eta|\leq 1/2$
\begin{equation}\label{hausdorff}
dist_{H}(\sigma(h_{\ve}),\sigma(h_{\eta})\leq K\frac{C}{\beta^{6}}|\ve -\eta|^{1/2}
\end{equation}
where $dist_{H}(A,B)$ is the Hausdorff distance between two compact sets in
$\mathbb{R}$. For the Harper operator (\ref{hausdorff}) has been proved in
\cite{avmsi} (improving an earlier result in \cite{ellchyu} giving the
exponent $1/3$ in (\ref{hausdorff})). For the more general case of magnetic
matrices (see (\ref{trans}), (\ref{const}))(\ref{hausdorff}) follows from the
existence of a 1/2-H\"{o}lder continuous field of rotation algebras
\cite{haro},\cite{kiph}. It seems \cite{be}, \cite{he} that the result in
Theorem 1i. is optimal in the sense that the exponent $1/2$ in (\ref{1/2})
cannot be improved {\em uniformly} in the gap's length. As concerning Theorem
1ii. and Theorem 2, for the case of magnetic matrices there exists a better
result due to Bellissard \cite{be}, namely that the gap boundaries are
actually Lipschitz continuous, i.e. the logarithmic factor in the r.h.s. of
(\ref{ln0}), (\ref{ln1}) can be removed. We believe this is true also in the
general case but we were not able (at least up to now) to prove it. Concerning
the dependence of constants upon gap's length, both the results in \cite{be}
and (\ref{ln1}) are far from optimal; one has to use Theorem 1i. for small
gaps and Theorem 2 for large ones.

The proofs in \cite{haro}, \cite{kiph}, \cite{be} (the proof in \cite{avmsi}
uses the specific form of the Harper operator) rest heavily on the fact that
under the conditions (\ref{trans}), (\ref{const}), $h_{\ve}$ belongs to a
rotation algebra and then one can use the powerfull techniques of
$C^{*}$-algebras theory. In other words the translation invariance,
(\ref{trans}), as well as the "homogeneity of the magnetic field"
,(\ref{const}), seems to be essential for these proofs.

The basic ideea of our proof is the one already used in \cite{nen1} (see also
\cite{nen2}) for the first general proof (see \cite{avsi} for a particular
case) of set theoretic continuity of the spectra of magnetic Schr\"{o}dinger
operators against variations of the magnetic field. It is based on exploiting,
at the technical level, the gauge symmetry. Accordingly, it is expected to
work under very general assumtions on $h(\bx, \by)$, and indeed while we 
restricted ourselvs to the discrete two-dimensional case, the results in
Theorems 1 and 2 can be generalized to cover higher dimensions, matrix (or
even operator) valued $h(\bx, \by)$ and more important, the continuous case
i.e. the case of "twisted" integral operators in $L^{2}(\mathbb{R}^{n},d\mu)$
\cite{nen4}.

Twisted integral operators in $L^{2}(\mathbb{R}^{n},d\mu)$ are intimately
related to magnetic Schr\"{o}dinger (and Dirac) operators. Let us outline, at
the heuristic level, the main ideea of this connection (see \cite{nen3} for
details). Let
\begin{equation}\label{5Hve}
H_{\ve,{\bf a}} = ({\bf P}-{\bf A}_{0}(\bx)-\ve {\bf a}(\bx))^{2}+V(\bx)
\end{equation}
with ${\bf b}(\bx)=curl {\bf a}(\bx)$ uniformly bounded together with its
first order derivatives and ${\bf A}_{0}(\bx)$, $V(\bx)$ satisfying the
appropriate conditions as to assure that  $H_{\ve}$  is a family of
semi-bounded self-adjoint operators in $L^{2}(\mathbb{R}^{n},d\bx),\;\;n=2,3$.
Take $-E_{0}$  sufficiently large so that $(-\infty, E_{0}+1)\subset
\rho(H_{\ve})$. Then it turns out that \cite{nen3}
 \begin{equation}\label{connection}
(H_{\ve}-E_{0})^{-1}=S_{\ve,E_{0}}+\ve V(\ve,E_{0})
\end{equation}
where $V(\ve,E_{0})$ is uniformly bounded as $\ve \rightarrow 0$ and $S_{\ve,E_{0}}$
is the integral operator:
\begin{equation}\label{Sa}
(S_{\ve, E_{0}}f)({\bf x})=
\int_{{\bf R}^{3}}
e^{i \ve \phi_{{\bf a}}({\bf x},{\bf y})}
G_{0}({\bf x},{\bf y};E_{0})f({\bf y})d{\bf y}
\end{equation}
where
\begin{equation}
\phi_{{\bf a}}({\bf x},{\bf y})=
 \int_{{\bf y}}^{{\bf x}}{\bf a}({\bf u})\cdot d{\bf u}
\end{equation}
and $G_{0}({\bf x},{\bf y};E_{0})$ is the integral kernel of 
$(({\bf P}-{\bf A}_{0}(\bx))^{2}+V(\bx)-E_{0})^{-1}$.
Notice that in this case (see(\ref{flux1})) $F(\bx,\by, \bz)$ is nothing but
the flux of ${\bf b}(\bx)$ through the triangle $\Delta(\bx,\by,\bz)$. Now the
spectral properties of $H_{\ve}$ can be read from the spectral properties of
$(H_{\ve}-E_{0})^{-1}$ and since the second term in the r.h.s. of (\ref{connection}) can
be controlled by regular perturbation theory one is led to the study of
$S_{\ve,E_{0}}$.
Along this way one obtains the analog of Theorems 1 and 2 for magnetic
Schr\"{o}dinger and (with an easy extension) Dirac operators. To our knowledge
the best result to date about set theoretic continuity of the spectra of
Schr\"{o}dinger and Dirac operators is the $2/3-\delta$-H\"{o}lder continuity
result in \cite{brco}(the $1/2$-H\"{o}lder continuity result is contained
implicitely in \cite{nen1} and \cite{hesj}).

On the way of proving Theorem 2ii. we obtain the following result about
smoothness of "almost" convex functions, which might be interesting in itself.
We give the result only in the one-dimensional case but  it can extended (as the similar
results for the mid-point convex functions \cite{arsu}) to a
more general context.
\begin{proposition}\label{conv}
Let $F:{\mathbb R}\rightarrow {\mathbb R}$ satisfying
\begin{equation}\label{interval}
\sup_{x}F(x)-\inf_{x}F(x) \leq 2P<\infty
\end{equation}
\begin{equation}
F(x)-\frac{F(x+\eta)+F(x-\eta)}{2} \leq N|\eta|^{\alpha};\;\;|\eta|\leq
1/2,\;x\in {\mathbb R}, \;N<\infty,\,\alpha >0.
\end{equation}
Then for $x,y\in {\mathbb R},\;\; |x-y|\leq1/2$:
\begin{equation}\label{fin}
|F(x)-F(y)|\leq \left\{\begin{array}{rcl}(4P+3N\frac{1}{1-2^{1-\alpha}})|x-y|&
\mbox{if}& \alpha >1\\  (4P+6N)|x-y||ln|x-y||& \mbox{if}& \alpha=1\\
(4P+2N\frac{1}{1-2^{\alpha -1}})|x-y|^{\alpha} & \mbox{if} & \alpha <1
\end{array}\right.  
\end{equation}
\end{proposition}

\section{The proofs.}
{\em Proof of Theorem 1}

It is sufficient to consider only the case $\eta=0$ : write
$h_{\ve}(\bx,\by)=h_{\eta}(\bx,\by)e^{i(\ve-\eta)\phi(\bx,\by)}$ and
observe that $h_{\eta}(\bx,\by)$ satisfies (\ref{herm}), (\ref{marg}).
Before entering the technicalities let us point out the main ideea of the
proof (borrowed fron \cite{nen1}). For ${\bf c}\in \bZ^{2}$, consider the
"gauge transformation"
$$
(U_{\bc,\ve}f)(\bx)=e^{i\ve \phi(\bx,\bc)}f(\bx).
$$
By direct computation for $E\in \bR$:
$$
U_{\bc,\ve}^{*}(h_{\ve}-E)= (h_{0}-E)U_{\bc,\ve}^{*}+
(U_{\bc,\ve}^{*}h_{\ve}U_{\bc,\ve}-h_{0})U_{\bc,\ve}^{*}.
$$
The main point is that (see the proof of (\ref{gauge} ) below) on functions,
$\psi$, supported on a ball centred at $\bc$ and of radius $L$,
$\parallel (U_{\bc,\ve}^{*}h_{\ve}U_{\bc,\ve}-h_{0})U_{\bc,\ve}^{*}
\psi\parallel $ is at most of order $L|\ve|\parallel \psi \parallel $. Suppose
now that $E\in \sigma (h_{\ve})$. Then one can find $\psi$ with $\parallel
\psi \parallel =1$ such that $\parallel (h_{\ve}-E)\psi \parallel $ is small.
If, in addition $\psi$ is localised somewhere, then one can find $\bc$ such that by the above argument 
$\parallel (h_{0}-E)U_{\bc,\ve}^{*}\psi \parallel $ is also small and then
$E$ must be close to $\sigma (h_{0})$. The trouble with this argument is that
the functions for which $\parallel (h_{\ve}-E)\psi \parallel $ is small might
not be localized. So one has either to localize them and estimate the
"localization error"(as done in \cite{avmsi} for the almost Mathieu operator)
or to design an appropriate, $\psi$ dependent "partition of unity" as done in
\cite{nen1}) We shall follow the second route.

A finite number of strictly positive, absolute constants will appear
during the proof; all of them will be denoted by $k>0$. Also a finite number of
finite, positive, absolute constants will appear and will be denoted by
$K <\infty$. We begin with a preliminary lemma containing a technical result.
It is patterned after a similar result in \cite{nen1}.
Let $\ba \in \bZ^{2},\;\;N\in \bN^{+}$,
\begin{equation}\label{cub}
C(\ba,N)=\{ \bx=(x_{1},x_{2})||x_{\mu}-a_{\mu}|\leq N,\;\;\mu=1,2\},
\end{equation}
$\chi_{\ba,N}$ the characteristic function of $C(\ba,N)$, $f_{N}(\bx)$
satisfying 
$$
0\leq f_{N}(\bx) \leq 1,
$$
\begin{equation}\label{f}
f_{N}(\bx)=\left\{\begin{array}{rcl}1& \mbox{if}& \bx \in C({\bf 0},N)\\ 
0& \mbox{if}& \bx \notin C({\bf 0},2N) \end{array}\right.
\end{equation}
$$
|f_{N}(\bx)-f_{N}(\by)|\leq \frac{|\bx-\by|}{N}
$$
and $f_{N,\ba}$ defined by:
\begin{equation}
f_{N,\ba}(\bx)=f_{N}(\bx-\ba).
\end{equation}
\begin{lemma}\label{partition}
Let $\Phi\in L^{2}(\bZ^{2})$ supported on a finite set, $N\in \bN^{+}$. Then
there exist $\ba_{0},\ba_{1},...,\ba_{p},\; p<\infty$ depending upon $\Phi$
such that:

i.
\begin{equation}\label{dist}
\min_{j\neq l}|\ba_{j}-\ba_{l}|\geq 8N.
\end{equation}
For $j\neq l$
\begin{equation}\label{dist1}
\rho_{j,l;N}=\min_{\bx\in C(\ba_{j},2N), \by\in C(\ba_{l},2N)}|\bx-\by| \geq
2N.
\end{equation}
There exist $K <\infty$ such that
\begin{equation}\label{exp}
\max_{j}\sum_{l\neq j}e^{-\beta \rho_{j,l;N}} \leq\frac{K}{\beta^{2}}e^{-\beta
N}.
\end{equation}

ii. If
\begin{equation}\label{fphin}
f_{N,\Phi}=\sum_{j=0}^{p}f_{N,\ba_{j}}
\end{equation}
then
\begin{equation}\label{min}
0 \leq f_{N,\Phi} \leq 1;\;\;|f_{N,\Phi}(\bx)-f_{N,\Phi}(\by)| \leq
\frac{|\bx-\by|}{N},
\end{equation}
\begin{equation}\label{max}
\parallel \Phi f_{N,\Phi} \parallel \geq \frac{\parallel \Phi \parallel}{9}.
\end{equation}

\end{lemma}

{\em Proof.} Let $\ba_{0}$ be a point of maximum of $\parallel \Phi
\chi_{\ba,N}\parallel$ (as a function of $\ba$). Define
\begin{equation}
\Phi_{1}=\Phi(1-\chi_{\ba_{0},9N})
\end{equation}
and repeat the procedure by taking $\ba_{1}$  to be a point
of maximum for $\parallel \Phi_{1}
\chi_{\ba,N}\parallel$ , etc. Since $\Phi$ is supported on a finite set, at some
$p<\infty$,  (depending upon $\Phi$) $\Phi_{p}(1-\chi_{\ba_{p},9N})
\equiv 0$ and the procedure stops. Now (\ref{dist}) holds true by construction
and (\ref{dist1}) follows at once from (\ref{dist}). For (\ref{exp}), set a
lattice centred at $\ba_{j}$ with the lattice spacing equal to $5N$. Due to
(\ref{dist}) each lattice cell contains at most one $\ba_{l}$. Move each
$\ba_{l}$ to the corner of the cell (containing it) which is nearest to
$\ba_{j}$. By this operation the sum in (\ref{exp}) is increased. The
resulting sum is dominated by 
$$
2\sum_{{\bf q}\in \bZ^{2},{\bf q}\neq {\bf 0}}e^{-\beta N |{\bf q}|}
$$ which gives (\ref{exp}).
Further, (\ref{min}) is almost obvious. Indeed by (\ref{dist1}),
$f_{N,\ba_{j}}$ have disjoint supports so  $0 \leq f_{N,\Phi} \leq 1$. Suppose
$f_{N, \Phi}(\bx)\neq 0,\;\;f_{N, \Phi}(\by) \neq 0$. Since $f_{N,\ba_{j}}$
have disjoint supports, $f_{N,
\Phi}(\bx)-f_{N,\Phi}(\by)=f_{N,\ba_{j}}(\bx)-f_{N,\ba_{l}}(\by)$ for some $j$
and $l$. If $j=l$ (\ref{min}) follows from (\ref{f}) and if $j\neq l$ then
$|\bx-\by| \geq 2N$ and then $|f_{N, \Phi}(\bx)-f_{N,\Phi}(\by)|\leq
\frac{|\bx-\by|}{2N}$. Suppose now $f_{N, \Phi}(\bx)\neq 0,\;\;
f_{N,\Phi}(\by)=0$ Then for some $j$, $f_{N, \Phi}(\bx)=f_{N,\ba_{j}}(\by)$
and $\by \notin supp f_{N, \ba_{j}}$. It follows that $f_{N,
\Phi}(\bx)-f_{N,\Phi}(\by)=f_{N, \ba_{j}}(\bx)=f_{N,
\ba_{j}}(\bx)-f_{N,\ba_{j}}(\by)$ and again (\ref{min}) follows from
(\ref{f}). The case $f_{N, \Phi}(\bx)= 0,\;\; f_{N,\Phi}(\by)\neq 0$ is
similar and the case  $f_{N, \Phi}(\bx)= f_{N,\Phi}(\by) =0$ is trivial.

Consider now
\begin{equation}
\tilde \Phi_{j}=\Phi f_{N,\ba_{j}}.
\end{equation}
Since by construction $\Phi_{j}=\Phi \prod_{l=0}^{j-1}(1-\chi_{\ba_{l},9N})$
\begin{equation}\label{maj}
|\tilde \Phi_{j}(\bx)|\geq |\Phi_{j}(\bx)f_{N,\ba_{j}}(\bx)|
\end{equation}
Now by definition (see also (\ref{maj}) and remember that $f_{N,\ba_{j}}$ have
disjoint supports)
$$
\parallel \Phi_{j}\parallel^{2} =\parallel
\Phi_{j}\chi_{\ba_{j},9N}\parallel^{2}+\parallel \Phi_{j+1}\parallel^{2} 
$$ 
so that
$$
\parallel \Phi\parallel^{2} =\sum_{j=0}^{p} \parallel
\Phi_{j}\chi_{\ba_{j},9N}\parallel^{2} \leq 81 \sum_{j=0}^{p} \parallel
\Phi_{j}\chi_{\ba_{j},N}\parallel^{2} 
$$
$$
\leq 81 \sum_{j=0}^{p} \parallel
\Phi_{j}f_{N,\ba_{j}}\parallel^{2}\leq 81 \sum_{j=0}^{p} \parallel
\tilde{\Phi}_{j}\parallel^{2}= 
81 \parallel \Phi \sum_{j=0}^{p}f_{N,\ba_{j}}
\parallel^{2} =81 \parallel f_{N,\Phi}\Phi \parallel^{2}
$$
and the proof of Lemma \ref{partition} is finished.

For the sake of easy quotation we collect some simple facts in:
\begin{lemma}\label{estimari}

i. Let $A$ be the operator given by
\begin{equation}
(Ag)(\bx)=\sum_{\by \in \bZ^{2}}e^{-\beta |\bx-\by |/2}g(\by)
\end{equation}
Then
\begin{equation}\label{normA}
a\equiv \parallel A \parallel \leq \frac{K}{\beta^{2}}
\end{equation}

ii.
\begin{equation}\label{ineg0}
\sup_{x>0}x^{m}e^{-\alpha x}=(\frac{m}{\alpha})^{m }e^{-m};\;\;m,\alpha >0
\end{equation}

iii.
\begin{equation}\label{normah}
H\equiv \sup_{\ve \in \bR}\parallel h_{\ve} \parallel
\leq \sum_{\by \in \bZ^{2}}e^{-\beta |\by |} \leq K\frac{C}{\beta^{2}}
\end{equation}
\end{lemma}

{\em End of proof of Theorem 1i.}

Let $0<\delta \leq 3\parallel h_{\ve}\parallel \leq 3H$ and suppose $E\in
\sigma (h_{\ve})$. Then we can find $\Psi_{\delta},\;\; \parallel
\Psi_{\delta} \parallel=1$ and $\parallel (h_{\ve}-E)\Psi_{\delta}\parallel
\leq \delta/4$. Further, there exists $\Phi_{\delta}$ supported on a finite
set such that $\parallel \Psi_{\delta}-\Phi_{\delta}\parallel \leq
\frac{\delta}{8H}$. Then 
\begin{equation}\label{Phi}
\parallel \Phi_{\delta} \parallel \geq 5/8 ,\;\;
\parallel (h_{\ve}-E)\Phi_{\delta}\parallel
\leq \delta .
\end{equation}

Let ( we omit to write some indices) $f,\;\tilde{\Phi}_{j},\,
\rho_{j,l}$ as given by Lemma \ref{partition} applied to $\Phi_{\delta}$.
In what follows we shall use the same letter for a function on $\bZ^{2}$ and
for the corresponding multiplication operator. From (\ref{min}) and (\ref{Phi})
($[\cdot,\cdot]$ means the commutator)
$$
\delta \parallel \Phi_{\delta}\parallel \geq \parallel (h_{\ve}-E)\Phi_{\delta}\parallel
\geq \parallel f(h_{\ve}-E)\Phi_{\delta}\parallel
\geq \parallel (h_{\ve}-E)f\Phi_{\delta}\parallel -\parallel
[f,h_{\ve}]\Phi_{\delta}\parallel
$$ which implies
\begin{equation}\label{h-efphy}
\parallel (h_{\ve}-E)f\Phi_{\delta}\parallel \leq \delta \parallel \Phi_{\delta}\parallel
+\parallel
[f,h_{\ve}]\Phi_{\delta}\parallel.
\end{equation}
We first estimate from below the l.h.s. of (\ref{h-efphy}). The first
observation is that for $j\neq l$,
$<(h_{\ve}-E)\tilde{\Phi}_{j},(h_{\ve}-E)\tilde{\Phi}_{l}>$ is small for large
$N$. Indeed from Lemma \ref{partition}, Lemma \ref{estimari} and 
$|\by-\bz|\leq |\bx -\bz|+|\bx- \by|$:
$$
|<(h_{\ve}-E)\tilde{\Phi}_{j},(h_{\ve}-E)\tilde{\Phi}_{j}>|=|\sum_{\bx,\by,\bz}\overline{
e^{i\ve\phi(\bx,\by)}(h(\bx,\by)-E\delta _{\bx,\by})\tilde{\Phi}_{j}(\by)}
$$
\begin{equation}
e^{i\ve\phi(\bx,\bz)}(h(\bx,\bz)-E\delta _{\bx,\bz})\tilde{\Phi}_{l}(\bz)|
\leq (C+H)^{2}a^{2}e^{-\frac{\beta}{2}\rho_{j,l}}\parallel
\tilde{\Phi}_{j}\parallel \parallel \tilde{\Phi}_{l}\parallel.
\end{equation}
Further
$$
\parallel (h_{\ve}-E)f\Phi_{\delta}\parallel^{2}=
\sum_{j} \parallel (h_{\ve}-E)\tilde{\Phi}_{j}\parallel^{2}+
\sum_{j\neq l}<(h_{\ve}-E)\tilde{\Phi}_{j},(h_{\ve}-E)\tilde{\Phi}_{j}>
$$
\begin{equation}\label{htildephi}
\geq \sum_{j} \parallel (h_{\ve}-E)\tilde{\Phi}_{j}\parallel^{2}
-(C+H)^{2}a^{2}\sum_{j \neq l}e^{-\frac{\beta}{2}\rho_{j,l}}\parallel
\tilde{\Phi}_{j}\parallel \parallel \tilde{\Phi}_{l}\parallel.
\end{equation}
Viewing the last sum in (\ref{htildephi}) as a scalar product in $l^{2}(\bN)$
of $\parallel \tilde{\Phi}_{j}\parallel$ with $(B\parallel
\tilde{\Phi}\parallel )_{j}\equiv \sum_{j \neq
l}e^{-\frac{\beta}{2}\rho_{j,l}}\parallel \tilde{\Phi}_{l}\parallel$ and
estimating the norm of $B$ by the Schur test using (\ref{exp}) one obtains:
$$
\parallel (h_{\ve}-E)f\Phi_{\delta}\parallel^{2} \geq
$$
\begin{equation}\label{htildephi1}
\geq \sum_{j} \parallel (h_{\ve}-E)\tilde{\Phi}_{j}\parallel^{2}
-Ka^{2}(C+H)^{2}e^{-\frac{\beta}{2}N}\frac{1}{\beta^{2}}\parallel f\Phi_{\delta} \parallel^{2}.
\end{equation}
As said at the begining of the proof the main point is that, due to the fact
that $\tilde \Phi_{j}$ are localised around $a_{j}$, one can estimate from below the sum
in the r.h.s. of (\ref{htildephi1}) in terms of $h_{0}$. Indeed, let $U_{j,
\ve}$ be the unitary operator (gauge transformation) defined by:
\begin{equation}\label{U}
(U_{j,\ve}g)(\bx)=e^{i\ve\phi(\bx,\ba_{j})}g(\bx).
\end{equation}
then
$$
\parallel (h_{\ve}-E)\tilde{\Phi}_{j}\parallel=
\parallel U_{j,\ve}^{*}(h_{\ve}-E)\tilde{\Phi}_{j}\parallel=
\parallel (U_{j,\ve}^{*}h_{\ve}U_{j,\ve}-h_{0})U_{j,\ve}^{*}\tilde{\Phi}_{j}
+(h_{0}-E)U_{j,\ve}^{*}\tilde{\Phi}_{j}\parallel
$$
which (by the triangle inequality) gives:
$$ \parallel (h_{0}-E)U_{j,\ve}^{*}\tilde{\Phi}_{j}\parallel \leq
\parallel h_{\ve}-E)\tilde{\Phi}_{j}\parallel +
\parallel (U_{j,\ve}^{*}h_{\ve}U_{j,\ve}-h_{0})U_{j,\ve}^{*}\tilde{\Phi}_{j}
\parallel
$$
and then
$$
\sum_{j}\parallel (h_{0}-E)U_{j,\ve}^{*}\tilde{\Phi}_{j}\parallel^{2} \leq
$$
\begin{equation}\label{tri}
\leq
2\sum_{j}\parallel (h_{\ve}-E)\tilde{\Phi}_{j}\parallel^{2} +
2\sum_{j}\parallel (U_{j,\ve}^{*}h_{\ve}U_{j,\ve}-h_{0})U_{j,\ve}^{*}\tilde{\Phi}_{j}
\parallel^{2}.
\end{equation}
We estimate now from above the last sum in the r.h.s. of (\ref{tri}). The
crucial computation  is:
\begin{equation}
(U_{j,\ve}^{*}h_{\ve}U_{j,\ve}-h_{0})U_{j,\ve}^{*}\tilde{\Phi}_{j}=
\sum_{\by}(e^{i\ve F(\bx, \by,
\ba_{j})}-1)h(\bx,\by)(U_{j,\ve}^{*}\tilde{\Phi}_{j})(\by).
\end{equation}
Then using (\ref{ineg0}), (\ref{flux}) and the fact that $\by \in supp
\tilde{\Phi}_{j}$ implies $|\by-\ba_{j}|\leq 2\sqrt{2}N$ one obtains:
$$
|(U_{j,\ve}^{*}h_{\ve}U_{j,\ve}-h_{0})U_{j,\ve}^{*}\tilde{\Phi}_{j}(\bx)|
\leq K|\ve|N\frac{C}{\beta}(A|\tilde{\Phi}_{j}|)(\bx)
$$
whereof by Lemma \ref{estimari}:
\begin{equation}\label{gauge}
\parallel (U_{j,\ve}^{*}h_{\ve}U_{j,\ve}-h_{0})U_{j,\ve}^{*}\tilde{\Phi}_{j}
\parallel^{2}
\leq K(|\ve|N\frac{C}{\beta^{3}})^{2}\parallel\tilde{\Phi}_{j}\parallel^{2}.
\end{equation}
Putting together (\ref{max}), (\ref{h-efphy}), (\ref{htildephi1}), (\ref{tri})
and (\ref{gauge}) (remark in addition that $\sum_{j}\parallel
\tilde{\Phi}_{j}\parallel^{2}=\parallel f\Phi_{\delta}\parallel^{2}$) one has:
$$
\sum_{j}\parallel (h_{0}-E)U_{j,\ve}^{*}\tilde{\Phi}_{j}\parallel^{2} \leq
2(\delta \parallel \Phi_{\delta}\parallel + \parallel
[f,h_{\ve}]\Phi_{\delta}\parallel)^{2}+
$$
\begin{equation}\label{estim}
+K(\frac{(C+H)^{2}}{\beta^{2}}e^{-\beta
N/2}+\ve^{2}N^{2}\frac{c^{2}}{\beta^{6}})\parallel \Phi_{\delta}\parallel^{2}.
\end{equation}
The control of $\parallel
[f,h_{\ve}]\Phi_{\delta}\parallel$ is easy: from (\ref{min}), (\ref{marg}) and
Lemma \ref{estimari}
$$
|[f,h_{\ve}](\bx,\by)| \leq |f(\bx)-f(\by)||h(\bx,\by)| \leq
K\frac{1}{N\beta}e^{-\frac{\beta}{2}|\bx-\by|}
$$ 
and then by Lemma \ref{estimari}
\begin{equation}\label{estim1}
\parallel
[f,h_{\ve}]\Phi_{\delta}\parallel \leq K \frac{C}{N\beta^{3}}
\parallel
\Phi_{\delta}\parallel.
\end{equation}
Let
\begin{equation}
\hat{\Phi}_{\delta}=\sum_{j}U_{j,\ve}^{*}
\tilde{\Phi}_{j}.
\end{equation}
Observe that
\begin{equation}\label{tilde}
\parallel\hat{\Phi}_{\delta}\parallel^{2}= \sum_{j}\parallel
\tilde{\Phi}_{j}\parallel^{2}=
\parallel f\Phi_{\delta}\parallel
\end{equation}
By the same estimation as in the proof of (\ref{htildephi1}):
$$
\parallel (h_{0}-E)f\hat{\Phi}_{\delta}\parallel^{2} \leq
$$
\begin{equation}\label{htildephy2}
\leq \sum_{j} \parallel (h_{0}-E)U_{j,\ve}^{*}\tilde{\Phi}_{j}\parallel^{2}
+Ka^{2}(C+H)^{2}\frac{1}{\beta^{2}}e^{-\frac{\beta}{2}N}\parallel
\hat{\Phi}_{\delta} \parallel^{2}. 
\end{equation}
Putting the things together (see (\ref{estim}), (\ref{estim1}), (\ref{tilde}),
 (\ref{htildephy2})) and taking into account that
$e^{-\frac{\beta}{2}N} \leq K\frac{1}{\beta^{2}N^{2}}$ one obtains:
$$
\parallel (h_{0}-E)f\hat{\Phi}_{\delta}\parallel^{2} \leq
$$
\begin{equation}\label{htildephi3}
\{ 2(\delta+K\frac{C}{\beta^{3}N})^{2} +K\frac{C^{2}}{\beta^{12}N^{2}}+
K\ve^{2}N^{2}\frac{C^{2}}{\beta^{6}}\}
\parallel\hat{\Phi}_{\delta}\parallel^{2}.
\end{equation}
Choosing $N=[\frac{1}{|\ve|^{1/2}}]$ (here $[\cdot]$ means the integer part)
(remember that $0< \beta \leq 1$) one has from (\ref{htildephi3})
$$
\parallel (h_{0}-E)f\hat{\Phi}_{\delta}\parallel^{2} \leq
(4\delta^{2}+K|\ve |\frac{C^{2}}{\beta^{12}})
\parallel\hat{\Phi}_{\delta}\parallel^{2}.
$$
which finishes the proof since $\delta$ can be taken arbitrarily small and 
$\parallel\hat{\Phi}_{\delta}\parallel\geq 5/72$.

{\em Proof of Theorem 1.ii.} Theorem 1.ii follows from Proposition \ref{conv}
and the following Lemma
\begin{lemma}\label{aconv}
There exists $K<\infty$ such that for $|\ve|\leq 1/2$:
\begin{equation}\label{conv1}
E_{+}(\ve_{0})-\frac{E_{+}(\ve_{0}+\ve)+E_{+}(\ve_{0}-\ve)}{2}\leq
K\frac{C}{\beta^{4}}|\ve|
\end{equation}
\end{lemma}

{\em Remark.}  The method of proof of Lemma \ref{aconv} also gives for $|\ve-\ve_{0}| \leq 1/2$:
\begin{equation}\label{2/3}
|E_{+}(\ve_{0})-E_{+}(\ve)| \leq
K\frac{C}{\beta^{4}}|\ve-\ve_{0}|^{2/3}
\end{equation}

{\em Proof of Lemma \ref{aconv}}. As before it is sufficient to consider the
case $\ve_{0}=0$. We shall use the fact that for a self-adjoint operator, $A$,
$$
\sup \sigma (A)= \sup_{\parallel g\parallel=1}<g,Ag>.
$$
The main point there is that this will allow to replace the $1/N$ dependence
of the "localization error"in (\ref{estim1}) by a better one namely $1/N^{2}$.
Suppose $E \in \sigma(h_{0})$, and let ,$
\ba_{j},\; f_{N,\ba_{j}},\;\Phi_{\delta},\;f,\;\tilde{\Phi}_{j}$ as in the
proof of Theorem 1.i. The following localization identity goes back at least
to Agmon \cite{agmon} (see also \cite{hisi},\cite{nen3})
$$
<\tilde{\Phi}_{j},(h_{0}-E)\tilde{\Phi}_{j}>=
$$
\begin{equation}\label{agmon}
=Re<f_{N,\ba_{j}}\Phi_{\delta},f_{N,\ba_{j}}(h_{0}
-E)\Phi_{\delta}>-1/2<\Phi_{\delta},[f_{N,\ba_{j}},
[f_{N,\ba_{j}},h_{0}]]\Phi_{\delta}>.
\end{equation}
On the other hand (see (\ref{U}) for $U_{j,\ve}$) from the definition of 
$E_{+}(\ve)$:
$$
<\tilde{\Phi}_{j},(h_{0}-E)\tilde{\Phi}_{j}>=
<\tilde{\Phi}_{j},\{(h_{0}-U_{j,\ve}^{*}h_{\ve}U_{j,\ve})+
(U_{j,\ve}^{*}h_{\ve}U_{j,\ve}-E_{+}(\ve))+(E_{+}(\ve)-E)\}
\tilde{\Phi}_{j}>\leq
$$
\begin{equation}\label{tht}
\leq (E-E_{+}(\ve))\parallel \tilde{\Phi}_{j} \parallel^{2} +
<\tilde{\Phi}_{j}, ((h_{0}-U_{j,\ve}^{*}h_{\ve}U_{j,\ve})\tilde{\Phi}_{j}>
\end{equation}
From (\ref{agmon}) and (\ref{tht}) (remember that $\parallel f\Phi_{\delta}
\parallel^{2}=\sum_{j}\parallel \tilde{\Phi}_{j} \parallel^{2}$):
$$
(E_{+}(\ve)-E)\parallel f\Phi_{\delta} \parallel^{2} \leq
-\sum_{j}Re<f_{N,\ba_{j}}\Phi_{\delta},f_{N,\ba_{j}}(h_{0}
-E)\Phi_{\delta}>+
$$
\begin{equation}\label{e-e+}
+\frac{1}{2}\sum_{j}<\Phi_{\delta},[f_{N,\ba_{j}},
[f_{N,\ba_{j}},h_{0}]]\Phi_{\delta}>+\sum_{j}
<\tilde{\Phi}_{j}, ((h_{0}-U_{j,\ve}^{*}h_{\ve}U_{j,\ve})\tilde{\Phi}_{j}>.
\end{equation}
Writing (\ref{e-e+}) also for $-\ve$ and summing up one obtains:
$$
E-\frac{E_{+}(\ve)+E_{+}(-\ve)}{2})
\parallel f\Phi_{\delta}
\parallel^{2} 
\leq -\sum_{j}Re<f_{N,\ba_{j}}\Phi_{\delta},f_{N,\ba_{j}}(h_{0}
-E)\Phi_{\delta}>+
$$
\begin{equation}\label{e-e+1}
+\frac{1}{2}\sum_{j}<\Phi_{\delta},[f_{N,\ba_{j}},
[f_{N,\ba_{j}},h_{0}]]\Phi_{\delta}>+\frac{1}{2}\sum_{j}
<\tilde{\Phi}_{j},
(2h_{0}-U_{j,\ve}^{*}h_{\ve}U_{j,\ve}-U_{j,-\ve}^{*}h_{-\ve}U_{j,-\ve})\tilde{\Phi}_{j}>. 
\end{equation}
We are left with the problem of estimating the r.h.s. of (\ref{e-e+1}). Using
twice the Cauchy-Schwartz inequality:
$$
-\sum_{j}Re<f_{N,\ba_{j}}\Phi_{\delta},f_{N,\ba_{j}}(h_{0}
-E)\Phi_{\delta}> \leq 
\sum_{j} \parallel  f_{N,\ba_{j}}\Phi_{\delta}\parallel
\parallel  f_{N,\ba_{j}}(h_{0}-E)\Phi_{\delta}\parallel
$$
\begin{equation}\label{ineg}
(\sum_{j}\parallel \tilde{\Phi}_{j} \parallel^{2})^{1/2}(\sum_{j}
\parallel  f_{N,\ba_{j}}(h_{0}-E)\Phi_{\delta}\parallel^{2})^{1/2}
\end{equation}
Taking into account that $f_{N,\ba_{j}}$ have disjoint supports,
$$\sum_{j}f_{N,\ba_{j}}(\bx)^{2}=f(\bx)^{2},$$ which together with $0\leq f(\bx)
\leq 1$, (\ref{Phi}) and (\ref{max}) gives
\begin{equation}\label{h-e}
\sum_{j}
\parallel  f_{N,\ba_{j}}(h_{0}-E)\Phi_{\delta}\parallel^{2}
=\parallel  f(h_{0}-E)\Phi_{\delta}\parallel^{2}\leq 81 \delta^{2}
\parallel  f\Phi_{\delta}\parallel^{2}.
\end{equation}
Combining (\ref{ineg}) with (\ref{h-e}) one has
\begin{equation}\label{re}
-\sum_{j}Re<f_{N,\ba_{j}}\Phi_{\delta},f_{N,\ba_{j}}(h_{0}
-E)\Phi_{\delta}> \leq 
81 \delta^{2}
\parallel  f\Phi_{\delta}\parallel^{2}.
\end{equation}
Consider now the second term in the r.h.s. of (\ref{e-e+1}).
Observe that since $\max_{j}(f_{N,\ba_{j}}(\bx)-f_{N,\ba_{j}}(\by))^{2}
\leq \frac{\bx- \by|^{2}}{N^{2}}$ and that, for fixed $\bx$ and $\by$ at most
two terms in the sum $\sum_{j}(f_{N,\ba_{j}}(\bx)-f_{N,\ba_{j}}(\by))^{2}$ are
nonzero one has that
\begin{equation}\label{fn}
\sum_{j}(f_{N,\ba_{j}}(\bx)-f_{N,\ba_{j}}(\by))^{2} \leq 
2\frac{\bx- \by|^{2}}{N^{2}}.
\end{equation}
Now from Lemma \ref{partition}, Lemma \ref{estimari} and (\ref{fn})
$$
\sum_{j}<\Phi_{\delta},[f_{N,\ba_{j}},
[f_{N,\ba_{j}},h_{0}]]\Phi_{\delta}>
=\sum_{j,\bx,\by}\overline{\Phi_{\delta}(\bx)}((f_{N,\ba_{j}}(\bx)-f_{N,\ba_{j}}(\by))^{2}
h(\bx,\by)\Phi_{\delta}(\by)=
$$
$$
\sum_{\bx,\by}\overline{\Phi_{\delta}(\bx)}(\sum_{j}(f_{N,\ba_{j}}(\bx)-f_{N,\ba_{j}}(\by))^{2})
h(\bx,\by)\Phi_{\delta}(\by)\leq
\frac{2}{N^{2}}C
\sum_{\bx,\by}|\overline{\Phi_{\delta}(\bx)}||\bx-\by|^{2}
e^{-\beta |\bx -\by|}|\Phi_{\delta}(\by)| \leq
$$
\begin{equation}\label{com}
\leq K\frac{C}{N^{2}\beta^{2}}<|\Phi_{\delta}|,A|\Phi_{\delta}|> \leq
K\frac{C}{N^{2}\beta^{2}}\parallel f\Phi_{\delta}\parallel^{2}.
\end{equation}
A strightforward computation gives:
$$
\frac{1}{2} \sum_{j}
<\tilde{\Phi}_{j},
(2h_{0}-U_{j,\ve}^{*}h_{\ve}U_{j,\ve}-U_{j,-\ve}^{*}h_{-\ve}U_{j,-\ve})\tilde{\Phi}_{j}>
=
$$
\begin{equation}\label{cos}
=\sum_{j,\bx,\by}\overline{\Phi_{\delta}(\bx)}(1-cos\ve F(\bx,\by,\ba_{j}))
h(\bx,\by)\Phi_{\delta}(\by)
\end{equation}
Since for $\by \in supp \tilde{\Phi}_{j},\; |\by-\ba_{j}| \leq 2\sqrt{2} N$ and
using Lemma \ref{estimari} (remember also that $|F(\bx,\by,\ba_{j})| \leq area
\Delta (\bx,\by,\ba_{j}) \leq |\bx-\by||\by-\ba_{j}|/2$ one obtains from
(\ref{cos}) \begin{equation}\label{etal}
\sum_{j}
<\tilde{\Phi}_{j},
(2h_{0}-U_{j,\ve}^{*}h_{\ve}U_{j,\ve}-U_{j,-\ve}^{*}h_{-\ve}U_{j,-\ve})\tilde{\Phi}_{j}>
\leq K \ve^{2}N^{2}\frac{C}{\beta^{4}}\parallel f\Phi_{\delta}\parallel^{2}.
\end{equation}
Summing up  (\ref{re}), \ref{com}) and (\ref{etal}) one gets:
\begin{equation}\label{sconv}
E-\frac{E_{+}(\ve)+E_{+}(-\ve)}{2})
\parallel f\Phi_{\delta}
\parallel^{2} \leq K(\delta
+\frac{C}{N^{2}\beta^{4}}+\ve^{2}N^{2}\frac{C}{\beta^{4}}) \parallel
f\Phi_{\delta}\parallel^{2}. \end{equation}
Choosing again $N=[\ve^{-1/2}]$ and taking into account that (\ref{sconv})
holds true for all $E\in \sigma(h_{0})$ and $\delta $ can be arbitrarily small, one has
\begin{equation}
 E_{+}(0)-\frac{E_{+}(\ve)+E_{+}(-\ve)}{2})
\leq K\frac{C}{\beta^{4}}|\ve| 
\end{equation}
and the proof of Lemma \ref{aconv} is complete.

{\em Remark.}  The above estimations applied to (\ref{e-e+}) leads to
$E_{+}(0)-E_{+}(\ve)
\leq K\frac{C}{\beta^{4}}|\ve|^{2/3}$. By interchanging $0$ and $\ve$ one obtains
also $E_{+}(\ve)-E_{+}(0)
\leq K\frac{C}{\beta^{4}}|\ve|^{2/3}$ and then
$|E_{+}(\ve)-E_{+}(0)|
\leq K\frac{C}{\beta^{4}}|\ve|^{2/3}$.

{\em Proof of Proposition \ref{conv}.} 
Without restricting the generality one can replace (\ref{interval}) by
\begin{equation}\label{int1}
|F(x)|\leq P.
\end{equation}
We  give the proof for $\alpha=1$ (the case we need)  and leave the
details to the reader the details for $\alpha \neq 1$. 

For $x\in \bR$ consider the function
\begin{equation}\label{g}
g_{x}(u)=F(x+u)-F(x),\; u\in \bR.
\end{equation}
 One has (uniformly in $x$):
\begin{equation}\label{int2}
|g_{x}(u)| \leq 2P,
\end{equation}
\begin{equation}\label{0}
g_{x}(0)=0,
\end{equation}
\begin{equation}\label{aconv1}
g_{x}(u)-\frac{g_{x}(u+\eta)+g_{x}(u-\eta)}{2} \leq N|\eta|,\; |\eta|\leq 1/2.
\end{equation}
Every $u\in (0,1/2]$ can be written (uniquely) as
\begin{equation}\label{u}
u=\frac{a}{2^{n}};\; a\in (1/2,1], \; n\in \bN^{+}.
\end{equation}
Let $a\in (1/2,1]$ be fixed. Then from (\ref{0}), (\ref{aconv1}) with $u=\eta
=a/2$ one has 
\begin{equation}\label{a/2}
g_{x}(\frac{a}{2})
\leq \frac{g_{x}(a)}{2}+N\frac{a}{2}.
\end{equation}
Then by induction over $n$:
\begin{equation}\label{a/2n}
g_{x}(\frac{a}{2^{n}})
\leq \frac{g_{x}(a)}{2^{n}}+Nn\frac{a}{2^{n}}.
\end{equation}
Indeed taking $u=\eta =\frac{a}{2^{n}}$ in (\ref{aconv1}):
$$
g_{x}(\frac{a}{2^{n}})
\leq \frac{1}{2}g(\frac{a}{2^{n-1}})+N\frac{a}{2^{n}} \leq 
$$
\begin{equation}
\leq
\frac{1}{2}[
 \frac{g_{x}(a)}{2^{n-1}}+N(n-1)\frac{a}{2^{n-1}}]+
N\frac{a}{2^{n}}=
\frac{g_{x}(a)}{2^{n}}+Nn\frac{a}{2^{n}}.
\end{equation}
Take now $u\in (0,1/2]$ and write it as in (\ref{u}). Since $ln 2 \geq 1/2$ one
has
\begin{equation}
n\leq 2 |ln \;u|.
\end{equation}
Then from (\ref{int2}) and $|ln \;u|\geq 1/2$:
\begin{equation}\label{ln}
g_{x}(u)=g_{x}(u)-g_{x}(0) \leq \frac{g_{x}(a)}{a}u+2Nu|ln \;u|
\leq 2(2P+N)u|ln \;u|.
\end{equation}
In the same way for $u\in [-1/2,0)$:
\begin{equation}\label{ln-}
g_{x}(u)-g_{x}(0) 
\leq 2(2P+N)|u||ln \;u|.
\end{equation}
Let $v\in (0,1/2]$. From (\ref{aconv1}) with $u=0,\;\eta =v$:
\begin{equation}
g_{x}(-v)\geq -g_{x}(v)-2Nv
\end{equation}
and using (\ref{ln})
\begin{equation}\label{-u}
g_{x}(-v)\geq -2(2P+N)v |ln\;v| -\frac{2N}{|ln \;v|}v |ln\;v|
\geq -(4P+6N)v |ln\;v|.
\end{equation}
In the same way for $u\in (0,1/2]$
$$
g_{x}(u)\geq -(4P+6N)u |ln\;u|
$$
which together with (\ref{-u}), (\ref{ln}) and(\ref{ln-})
\begin{equation}\label{afin}
|g_{x}(u)-g_{x}(0) |=|g_{x}(u)|
\leq (4P+6N)u|ln \;u| ,\;\; |u\leq 1/2.
\end{equation}
Writing (\ref{afin}) in terms of $F(x)$ one obtains (\ref{fin}) for $\alpha=1$.

{\em Proof of Theorem 2.} Again it is sufficient to consider $\ve_{0}=0$.
We shall prove (\ref{ln1}) for $E_{1}(\ve)$; the proof for $E_{2}(\ve)$
is similar. In what follows a finite number of constants depending upon $C$
and $\beta $ will appear; they are all denoted $m>0$ (when we want to stress
that they are strictly positive) of $M<\infty$ (when we want to stress that
they are positive and finite). Let $\Gamma_{j}$ be contours of finite length
enclosing $\sigma_{j}(\ve)$ such that
$ \sup_{z \in \Gamma_{1} \cup \Gamma_{2}, |\ve|\leq md^{2}} dist(z, \sigma
(h_{\ve}) \geq d$. Let $\lambda < \inf_{\ve}\sigma (h_{\ve})$ and consider for 
$|\ve|\leq md$:
\begin{equation}\label{h1}
h_{1}(\ve)= \frac{i}{2\pi}\int_{\Gamma_{1}}z(h_{\ve}-z)^{-1}dz +
(\lambda -1)\frac{i}{2\pi}\int_{\Gamma_{2}}(h_{\ve}-z)^{-1}dz.
\end{equation}
By construction (for $|\ve|\leq md^{2}$)
\begin{equation}\label{E1}
E_{1}(\ve)=\sup \sigma(h_{1}(\ve)).
\end{equation}
We shall prove that up to errors which are Lipschitz (in norm), $h_{1}(\ve)$
has the same form as $h_{\ve}$ (with a $d$ dependent $\beta$!).
We begin by estimating $\sup_{z\in \Gamma_{1}\cup \Gamma_{2}}|G_{0}(\bx,
\by;z)|$ where $G_{0}(\bx,
\by;z)\equiv (h_{0}-z)^{-1}(\bx,\by)$. For that we use (like in \cite{nen4})
some elementary facts from Agmon-Combes-Thomas theory (see e.g. \cite{hisi}).
Consider for $\mu\in \bR^{+},\;\bx_{0}\in \bZ^{2}$ the rotated operator
\begin{equation}\label{hrot}
h_{\mu,\bx_{0}}=e^{\mu|\cdot-\bx_{0}|}h_{0}e^{-\mu|\cdot-\bx_{0}|}.
\end{equation}
given by the kernel
\begin{equation}\label{hrotk}
h_{\mu,\bx_{0}}(\bx,\by)=e^{\mu|\bx-\bx_{0}|}h(\bx,\by)e^{-\mu|\by-\bx_{0}|}.
\end{equation}
Using $||\bx-\bx_{0}|-|\by-\bx_{0}|| \leq |\bx-\by|;\;\;|e^{x}-1| \leq
|x|e^{|x|}$ one has
\begin{equation}\label{ex-1}
|e^{\mu (|\bx-\bx_{0}|-|\by-\bx_{0}|)}-1| \leq \mu |\bx-\by|e^{\mu|\bx-\by|}. 
\end{equation}
If one writes
$$
h_{\mu,\bx_{0}}=h_{0}+\mu B_{\mu,\bx_{0}}
$$ 
then from (\ref{hrotk}) and (\ref{ex-1})
\begin{equation}\label{Bb}
\sup_{\bx_{0}\in \bZ^{2}, 0\leq \mu \leq \beta/2} \parallel B_{\mu,\bx_{0}}
\parallel \equiv b <\infty.
\end{equation}
Since $\sup_{z\in \Gamma_{1}\cup \Gamma_{2}}\parallel (h_{0}-z)^{-1} \parallel
\leq \frac{1}{d}$, for
\begin{equation}\label{mu}
\mu \leq min \{ \frac{\beta}{2},\frac{d}{2b} \}
\end{equation}
and $z\in \Gamma_{1}\cup \Gamma_{2}$:
$$
\mu \parallel B_{\mu,\bx_{0}}(h_{0}-z)^{-1} \parallel \leq 1/2
$$
so that by perturbation theory, on $\Gamma_{1}\cup \Gamma_{2}$
$$
(h_{\mu,\bx_{0}}-z)^{-1}=(h_{0}-z)^{-1}[1+\mu
B_{\mu,\bx_{0}}(h_{0}-z)^{-1}]^{-1}
$$ 
and
\begin{equation}\label{bhrot}
\sup_{z\in \Gamma_{1}\cup \Gamma_{2}}\parallel (h_{\mu,\bx_{0}}-z)^{-1}.
\parallel \leq \frac{2}{d}
\end{equation}
Due to the fact that $e^{-\mu|\cdot-\bx_{0}|}$ is injective, 
$(h_{0}-z)Range (e^{-\mu|\cdot-\bx_{0}|} )\subset Range
(e^{-\mu|\cdot-\bx_{0}|})$ and that $h_{\mu,\bx_{0}}-z$ has a bounded inverse
one has that
$(h_{0}-z)^{-1}Range (e^{-\mu|\cdot-\bx_{0}|} )\subset Range
(e^{-\mu|\cdot-\bx_{0}|})$  and
\begin{equation}\label{rezrot}
e^{\mu|\cdot-\bx_{0}|}(h_{0}-z)^{-1}e^{-\mu|\cdot-\bx_{0}|}=
(h_{\mu,\bx_{0}}-z)^{-1}.
\end{equation}
Let now for $\bc \in \bZ^{2}$
$$
\Psi_{\bc}(\bx)=\delta_{\bx,\bc}
$$
where $\delta_{\bx,\bc}$ is the usual Kronecker symbol.
Then by (\ref{bhrot}) and (\ref{rezrot})
$$
\sup_{z\in \Gamma_{1}\cup \Gamma_{2}}|G_{0}(\bx,
\by;z)|=
\sup_{z\in \Gamma_{1}\cup \Gamma_{2}}|<\Psi_{\bx},(h_{0}-z)^{-1}\Psi_{\by}>|=
$$
\begin{equation}\label{erez}
=\sup_{z\in \Gamma_{1}\cup \Gamma_{2}}|<e^{-\mu|\cdot-\by|}\Psi_{\bx}, (h_{\mu,\bx_{0}}-z)^{-1}
e^{\mu|\cdot-\by|}\Psi_{\by}>|
\leq \frac{2}{d}e^{-\mu|\bx-\by|}.
\end{equation}
Let now $S_{\ve,z}$ be the operator given by
\begin{equation}\label{sz}
(S_{\ve,z}g)(\bx)=\sum_{\by}e^{i\ve \phi (\bx,\by)}G_{0}(\bx,\by;z)g(\by).
\end{equation}
By direct computation
\begin{equation}
(h_{\ve}-z)S_{\ve,z}=1+\ve T_{\ve,z}
\end{equation}
with
\begin{equation}\label{tz}
T_{\ve,z}(\bx,\by)=e^{i\ve \phi (\bx,\by)}
\sum_{\bu}\frac{e^{i\ve F (\bx,\bu,\by)}-1}{\ve}(h(\bx,\bu-z\delta_{\bx,\bu})
G_{0}(\bu,\by;z).
\end{equation}
From (\ref{marg}), (\ref{flux}) and (\ref{erez}), for $z \in \Gamma_{1}\cup
\Gamma_{2}$:
$$
|T_{\ve,z}(\bx,\by)| \leq \frac{M}{d}
\sum_{\bu}|\bx-\bu||\by-\bu|e^{-\beta|\bx-\bu|}e^{-\mu|\by-\bu|}
$$
and then from the Young inequality
\begin{equation}\label{normt}
\sup_{\by}\sum_{\bx}|T_{\ve,z}(\bx,\by)| \leq \frac{M}{\mu^{3} d}.
\end{equation}
For
\begin{equation}
|\ve| \leq m \mu^{3}d
\end{equation}
\begin{equation}
\ve\parallel T_{\ve,z}\parallel \leq 1/2
\end{equation}
and then for $z \in \Gamma_{1}\cup
\Gamma_{2}$:
\begin{equation}\label{echiv}
(h_{\ve}-z)^{-1}=S_{\ve,z}(1+\ve T_{\ve,z})^{-1}=
S_{\ve,z}
-\ve S_{\ve,z}
T_{\ve,z}
(1+\ve T_{\ve,z})^{-1}\equiv
S_{\ve,z}+
\ve V_{\ve,z}
\end{equation}
with 
\begin{equation}
\parallel V_{\ve,z}\parallel \leq 2\parallel S_{\ve,z}\parallel
\parallel T_{\ve,z}\parallel \leq \frac{M}{d^{2}\mu^{5}}.
\end{equation}
By the definition of $h_{1}(\ve)$ (see (\ref{h1}) and (\ref{echiv})
\begin{equation}\label{tildeh}
h_{1}(\ve)=\tilde{ h}_{\ve}+\ve W_{\ve}
\end{equation}
where
\begin{equation}
\tilde{ h}_{\ve}=
\frac{i}{2\pi}\int_{\Gamma_{1}}zS_{\ve,z}dz +
(\lambda -1)\frac{i}{2\pi}\int_{\Gamma_{2}}S_{\ve,z}dz.
\end{equation}
From (\ref{sz})
\begin{equation}
\tilde{ h}_{\ve}(\bx,\by)=e^{i\ve \phi (\bx,\by)}\tilde{ h}(\bx,\by);
|\tilde{ h}(\bx,\by)| \leq \frac{M}{d}e^{-\mu|\bx-\by|}.
\end{equation} 
Take now (see (\ref{mu})) $\mu=md$. Since $W_{\ve}$ is uniformly bounded as
$\ve \rightarrow 0$ by perturbation theory
$ | \sup \sigma(h_{1}(\ve)-\sup \sigma (\tilde h_{\ve})| \leq Md^{-7}$, and the
application of Theorem 1.ii to $\tilde h_{\ve}$ finishes the proof of Theorem
2.

{\bf Acknowledgements.}

The paper has been partly written at Mittag-Leffler Institute during a one 
month participation in the programme "{\em
Partial differential equations and spectral theory}", Fall 2002. 
I thank the scientific steering committee for the invitation and Mittag-Leffler Institute for financial support.

    \end{document}